\documentclass[12pt,twoside]{article}
\usepackage[mathscr]{eucal}
\usepackage{amsmath,amsfonts,amssymb,amsthm,mathabx,empheq}
\usepackage{times}
\usepackage{pdfsync}
\usepackage{cite}
\usepackage{url}
\usepackage{hyperref}
\usepackage{tensor}
\usepackage{color}
\usepackage{multicol}
\usepackage{bbold}

\advance\textheight\topskip
\allowdisplaybreaks[1]

\newcommand\td{\text{d}}

\newcommand{\p}{\partial}

\newcommand{\be}{\begin{equation}}
\newcommand{\ee}{\end{equation}}
\newcommand{\bea}{\begin{eqnarray}}
\newcommand{\eea}{\end{eqnarray}}

\def\n{\nabla}
\def\bm{\bar{m}}

\newcommand{\nn}{\nonumber}
\newcommand*\xbar[1]{%
  \hbox{%
    \vbox{%
      \hrule height 0.5pt % The actual bar
      \kern0.3ex%         % Distance between bar and symbol
      \hbox{%
        \kern-0.0em%      % Shortening on the left side
        \ensuremath{#1}%
        \kern-0.0em%      % Shortening on the right side
      }%
    }%
  }%
}

\allowdisplaybreaks[1]

\DeclareFontFamily{OT1}{rsfs}{} \DeclareFontShape{OT1}{rsfs}{m}{n}{
<-7> rsfs5 <7-10> rsfs7 <10-> rsfs10}{}
\DeclareMathAlphabet{\mycal}{OT1}{rsfs}{m}{n}

\hypersetup{
colorlinks=true,
linktoc=page,    %set to all if you want both sections and subsections linked
linkcolor=blue,
citecolor=blue,
urlcolor=blue,%magenta,
colorlinks=true,
}

\begin{document}

\title{Five dimensional Weyl double copy}

\author{Weicheng Zhao, Pu-Jian Mao, and Jun-Bao Wu}

\date{}

\def\mytitle{Five dimensional Weyl double copy}

\begin{flushright}
\tt USTC-ICTS/PCFT-24-30
\end{flushright}

\addtolength{\headsep}{4pt}

\begin{centering}

  \vspace{1cm}

  \textbf{\large{\mytitle}}

  \vspace{1.5cm}

  {\large Weicheng Zhao$^a\,$\footnote{
The unusual ordering of authors instead of the standard alphabetical one in hep-th community is for
students to get proper recognition of contribution under the current out-dated practice in China.
}, Pu-Jian Mao$^{a}$, and Jun-Bao Wu$^{a,b}$ }

\vspace{0.5cm}

\begin{minipage}{.9\textwidth}\small \it  \begin{center}
     ${}^{a}$ Center for Joint Quantum Studies and Department of Physics,\\
     School of Science, Tianjin University, 135 Yaguan Road, Tianjin 300350, China
 \end{center}
\end{minipage}

\vspace{0.3cm}

\begin{minipage}{.9\textwidth}\small \it  \begin{center}
    ${}^{b}$ Peng Huanwu Center for Fundamental Theory,  Hefei, Anhui 230026, China
 \end{center}
 \end{minipage}

\vspace{0.3cm}

\end{centering}

\begin{center}
Emails: zhaoweichengok@tju.edu.cn,\, pjmao@tju.edu.cn,\, junbao.wu@tju.edu.cn
\end{center}

\begin{center}
\begin{minipage}{.9\textwidth}
\textsc{Abstract}: The Weyl double copy (WDC) relation connects the Weyl tensor of the gravity theory
and the field strength tensor of the Maxwell theory, which provides a concrete realization of the classical double copy. Although intensively investigated, the WDC is only limited in four-dimensional spacetime. In this Letter, we generalize the WDC relation to five-dimensional spacetime, which offers the first example of the WDC in higher dimensions. We show that a special class of five-dimensional type N vacuum solutions admits a special class of degenerate Maxwell field that squares to give the Weyl tensor. The five-dimensional WDC relation defines a scalar field that satisfies the source-free Klein-Gordon equation on the curved background. We further verify that for five-dimensional pp-wave solution and Kundt solutions, the Maxwell fields and the scalar fields also satisfy the Maxwell's equations and the wave equation on five-dimensional Minkowski spacetime.

\end{minipage}
\end{center}

\thispagestyle{empty}
%\newpage

\section{Introduction}

Gauge and gravity theories are fundamentally important in our understanding of physical phenomena. It is intriguing to observe that they are intimately linked. One of such example is the celebrated AdS/CFT correspondence \cite{Maldacena:1997re}, which relates the quantum gravity in the anti-de Sitter (AdS) space to the conformal field theory (CFT) on the AdS boundary. Another striking example is the double copy relation, which interprets the perturbative scattering amplitudes of gravity as a product of two scattering amplitudes of gauge theory \cite{Kawai:1985xq,Bern:2008qj,Bern:2010ue}. The double copy perspective suggests a more efficient and elegant way for computing the gravity scattering amplitudes than the traditional calculations in General Relativity, see, e.g., \cite{Bern:2019prr,Bern:2022wqg,White:2017mwc,Adamo:2022dcm}, which plays a crucial role in the recently established amplitudes-based methods to derive the state-of-the-art results of interest to the gravitational wave community  \cite{Bern:2019nnu,Bern:2019crd,Bern:2021dqo,Bern:2021yeh}.

The remarkable success of the double copy relation has motivated investigations at the classical level connecting classical solutions of gauge and gravity theories which pioneers a classical double copy relation from the Kerr-Schild construction \cite{Monteiro:2014cda}. The classical double copy reveals exact relations between solutions beyond perturbative approach, which has a broader range of interests outside of the amplitudes community and significantly extends the scope of the double copy relation, see, e.g., \cite{Luna:2015paa,Luna:2016due,Goldberger:2016iau,Goldberger:2017frp,Bahjat-Abbas:2017htu,Carrillo-Gonzalez:2017iyj,Shen:2018ebu,Lee:2018gxc,Berman:2018hwd,CarrilloGonzalez:2019gof,Easson:2020esh,Gumus:2020hbb,Almeida:2020mrg,Campiglia:2021srh,Alkac:2021bav,Alkac:2021seh,Chacon:2021hfe,Adamo:2021dfg,Mao:2021kxq,Shi:2021qsb,Didenko:2022qxq,Chawla:2023bsu,Ceresole:2023wxg,Ferrero:2024eva}. Shortly, a second version of classical double copy, the Weyl double copy \cite{Luna:2018dpt}, was proposed by connecting the Weyl tensor of the gravity theory and the field strength tensor of the Maxwell theory.

The advantage of the WDC is resided in its gauge invariant nature, which provides a coordinate independent, hence more general method for revealing the exact mathematical connections between two classical theories. The WDC has drawn intensive attentions from various perspectives \cite{Keeler:2020rcv,Alawadhi:2020jrv,Godazgar:2020zbv,White:2020sfn,Monteiro:2020plf,Chacon:2021wbr,Godazgar:2021iae,Easson:2021asd,Chacon:2021lox,Han:2022ubu,Han:2022mze,Luna:2022dxo,Easson:2022zoh,Easson:2023dbk,Alkac:2023glx,Mao:2023yle,Liu:2024byr,Chawla:2024mse,Armstrong-Williams:2024bog}, which underpins the intrinsic connection between the Einstein equations and the Maxwell's equations. Although, the WDC was expected to exist in higher dimensions since it was proposed \cite{Luna:2018dpt} and the higher-dimensional spinorial formalism was adapted in view of this goal \cite{Monteiro:2018xev}, the current investigations are still restricted in four-dimensional spacetime.\footnote{In principle, the WDC relation \eqref{WDC} can be applied for particular exact solutions in any dimension if there is a spinorial description of the relevant fields, e.g., the single copy of the Kerr-NUT-(A)dS spacetime in five dimensions was derived from the WDC relation \cite{Chawla:2022ogv}. In \cite{Alawadhi:2020jrv}, the authors examine the purely algebraic Weyl doubling constraints to the Weyl
  tensor in generic dimensions, see also \cite{Didenko:2011ir} earlier ideas in
  this direction. In three-dimensional spacetime, a new
version of classical double copy has been proposed in \cite{CarrilloGonzalez:2022mxx,CarrilloGonzalez:2022ggn}, which involves the
  Cotton tensor rather than the Weyl tensor.} Since the independent components of the Weyl tensor increase much faster than the metric as the spacetime dimension increases, building exact relations for the Weyl tensor to realize a classical double copy in higher dimensions is much more challenging than the Kerr-Schild construction, see, e.g., the known examples of the later \cite{Carrillo-Gonzalez:2017iyj,Alkac:2021bav,Ortaggio:2023rzp,Ortaggio:2023cdz}. If the WDC relation exists in higher dimensions is a vital issue for its scope, origin \cite{White:2020sfn,Chacon:2021wbr}, and compatibility with the Kerr-Schild double copy \cite{Easson:2023dbk,Alkac:2023glx}. More importantly, it restricts how general the classical relation between gauge and gravity theories is.

In this Letter, we initialize the study of a five-dimensional Weyl double copy. The construction of the four-dimensional WDC formula constricts that it is only valid for the algebraic type D and type N spacetime.\footnote{Here, we consider two equal electromagnetic fields. The WDC relation can be defined in other types of spacetime for two different electromagnetic fields \cite{White:2020sfn}.} We start from the simpler case in five dimensions, the type N vacuum solution of Einstein gravity in the Coley-Milson-Pravda-Pravdova (CMPP) classification \cite{Coley:2004jv,Milson:2004jx,Pravda:2004ka,Ortaggio:2012jd}. In the five-dimensional spinorial formalism \cite{Monteiro:2018xev}, we propose an algebraic construction of the WDC formula for type N solutions. We find a self-contained reduction of the type N spacetime, where one can confirm that any solution of this special class admits a special class of degenerate Maxwell field that squares to give the Weyl tensor. Moreover, a complex scalar field is defined from this five-dimensional WDC relation and the scalar field satisfies the wave equation on the curved background. Hence, a concrete WDC relation is uncovered in five dimensions. We then present two examples of exact solutions in five dimensions, which are a special class of the pp-wave and Kundt solutions. Remarkably, the Maxwell fields and the scalar fields for those two cases also satisfy the Maxwell’s equations and the wave equation on the flat Minkowski spacetime. Our results confirm the existence of the higher-dimensional WDC and consolidate the robustness of the classical double copy relation.

%%%%%%%%%%%%%%%%%%%%%%%%%%%%%%%%%%%%%%%%%%%%%%%%%%%%%%%%%%%%%%%%%%%%%%%%%%%%%%%%%%%%%%%%%%%%%%%%%%%%%%%%%%%%%%%%%%%%%%%%%%%%%%%%%%%%%%%%%%%%%%%%%%%%%%%%%%%%%%%%%%%%%%%%%%%%%%%%%%%%%%%%%%%%%%%%%%%%%%%%%%%%%%%%%%%%%%%%%%%%%%%%%%%%%%%%%%%%%%%%

\section{4d WDC relation for type N solutions}

The WDC relation is better appreciated in the spinorial formalism \cite{Luna:2018dpt,Godazgar:2020zbv}. In four-dimensional spacetime, the homomorphism between the Lorentz group and SL(2,$\mathbb{C}$) allows one to convert the spacetime indices $\mu,\nu...$ into spinor indices $A,B,...$ and their conjugate $A',B',...$, where the Van der Waerden matrices $\sigma^\mu_{A A'}$ are applied to transform between them. The spinorial version of the Weyl tensor is fully determined by the totally symmetric Weyl spinor $\psi_{ABCD}$ and its complex conjugate. The WDC relation is interpreted by the decomposition of the Weyl spinor \cite{Luna:2018dpt}
\begin{equation}\label{WDC}
    \psi_{ABCD}=\frac{3c}{S}\phi_{(AB}\phi_{CD)},
\end{equation}
where $\phi_{AB}$ is a totally symmetric two-spinor defined from the Maxwell tensor $F_{\mu\nu}$ and $S$ is a (complex) scalar field. The WDC formula in \eqref{WDC} can be written in a null frame system, such as the Newman-Penrose (NP) formalism \cite{Newman:1961qr}. The null bases $(l,n,m,\bm)$ of the NP formalism is constructed from the spinor bases $\{o_A,\iota_A\}$ as
\be
\begin{split}
&l^\mu \sim \sigma^\mu_{A A'} o^A\bar{o}^{A'},\quad n^\mu \sim \sigma^\mu_{A A'} \iota^A \bar{\iota}^{A'},\\
&m^\mu \sim \sigma^\mu_{A A'} o^A \bar{\iota}^{A'},\quad \bm^\mu \sim \sigma^\mu_{A A'} \iota^A \bar{o}^{A'}.
\end{split}
\ee
The spinor indices are raised and lowered by the two-dimensional Levi-Civita tensor which can be decomposed by the spinor bases as $\epsilon_{AB}=o_A \iota_B-o_B\iota_A$.

For a type N spacetime, the only non-zero Weyl scalar is $\Psi_4=\psi_{ABCD}\iota^A \iota^B \iota^C\iota^D=C_{\mu\nu\rho\sigma}n^\mu \bm^\nu n^\rho \bm^\sigma$ when $l$ is the principal null direction. And $\Phi_2=\phi_{AB}\iota^A \iota^B=F_{\mu\nu}\bm^\mu n^\nu$ is the only non-zero Maxwell scalar of a degenerate Maxwell field. The type N WDC relation in the NP formalism is simply given by \cite{Godazgar:2020zbv}
\be\label{wdcnp}
\Psi_4=\frac{3c}{S}\Phi_2\Phi_2.
\ee
This relation seems revealing a trivial connection, because for any given Maxwell field, its square is naturally associated to the Weyl tensor from the algebraic derivation of the scalar field $S= 3c \Phi_2\Phi_2/\Psi_4$. The key point of the WDC is that if one is given a type N spacetime, how can one construct the Maxwell field which gives the Weyl tensor from its square, and what is the property of the scalar field. Those questions are well addressed in the seminal work \cite{Godazgar:2020zbv}. Inserting the WDC relation into the Bianchi identity of the Weyl tensor and assuming that the Maxwell scalar solves the Maxwell's equation, one can obtain the equations for the scalar field $S$ as
\be\label{S4}
D\log S - \rho=0,\quad \delta \log S - \tau=0.
\ee
If a scalar field is a solution of \eqref{S4}, it must solve the wave equation $\Box S=0$ on the type N background \cite{Godazgar:2020zbv}. For any scalar field solves \eqref{S4}, a degenerate Maxwell field can be derived from the WDC relation \eqref{wdcnp} as the square-root $\sqrt{S\Psi_4/(3c)}$. The integrability of equations in \eqref{S4} guarantees that all four-dimensional type N vacuum solutions admit a degenerate Maxwell field that squares to give the Weyl tensor and the zeroth copy (the scalar field $S$) solves the wave equation. The decomposition of the Weyl tensor is not unique in this case.

The aim of the present work is to extend the above analysis to five dimensions. However, the four-dimensional WDC relation can not imply any relation in five dimensions. The Weyl spinor in four dimensions can always be decomposed in terms of four rank-one spinors, which provides an alternative viewpoint of the Petrov classification from the alignment of the rank-one spinors. Hence, the four-dimensional WDC can only be constructed for algebraic type D and type N solutions for two equal Maxwell fields, which implies the intrinsic relationship between the four-dimensional WDC and the algebraic classification, and also simplifies the verification of the WDC, namely one just needs to consider two types of algebraically special vacuum solutions. In five dimensions, the algebraic classification is only relevant to the little group four-spinors of the Weyl tensor rather than the Weyl spinor \cite{Monteiro:2018xev}. If one insists on the relation \eqref{WDC} in five dimensions, the algebraic classification would not be of benefit to the verification of the WDC at all. A relevant issue is that there are more components of the Weyl tensor in five dimensions which makes the proposal of the WDC much more challenging. The very first issue about a five-dimensional WDC relation is that it should involve the Weyl spinor or the little group spinors. Actually, the existence of the WDC in five dimensions is also questionable. The four-dimensional WDC is argued to be originated from the twistor space \cite{White:2020sfn}. Though it has higher-dimensional generalization, the twistor space has less direct connection to spacetime in higher dimensions. So, it is very doubtable that a higher-dimensional twistor space can yield a higher-dimensional WDC that indicates a decomposition of the spacetime Weyl tensor.

%%%%%%%%%%%%%%%%%%%%%%%%%%%%%%%%%%%%%%%%%%%%%%%%%%%%%%%%%%%%%%%%%%%%%%%%%%%%%%%%%%%%%%%%%%%%%%%%%%%%%%%%%%%%%%%%%%%%%%%%%%%%%%%%%%%%%%%%%%%%%%%%%%%%%%%%%%%%%%%%%%%%%%%%%%%%%%%%%%%%%%%%%%%%%%%%%%%%%%%%%%%%%%%%%%%%%%%%%%%%%%%%%%%%%%%%%%%%%%%%

\section{5d WDC relation for type N solutions}

We follow closely \cite{Monteiro:2018xev} for the spinorial formalism in five dimensions. Five $\gamma$ matrices are chosen as the bases to connect the spacetime indices and spinor indices. They are given by
\be
\gamma^{\hat \mu}_{AB}=\begin{pmatrix}
0 & {\sigma^{\hat \mu \alpha}}_{\beta'}\\
\tilde{\sigma}^{{\hat \mu} \,\,\beta}_{\,\,\alpha'} & 0
\end{pmatrix},\quad
 \gamma^{4}_{A B}=- i \begin{pmatrix}
\epsilon^{\alpha \beta} & 0\\
0 & \epsilon_{\alpha' \beta'}
\end{pmatrix},
\ee
where $ \hat{\mu}=0,1,2,3$ denotes the first four components of a five-dimensional vector, $A,B=1,2,3,4$ are the spacetime spinor indices and $\alpha,\beta$ are now the spinor indices in the Van der Waerden matrices. Correspondingly, the spinor bases in five dimensions are chosen as
\be
\begin{split}
&k^A_1=\begin{pmatrix}
0 \\
\bar{o}^{\alpha'}
\end{pmatrix},\quad
k^A_2=\begin{pmatrix}
 o_\alpha\\
 0
\end{pmatrix},\\
&n^A_1=\begin{pmatrix}
\iota_\alpha  \\
0 
\end{pmatrix},\quad
n^A_2=-\begin{pmatrix}
0\\
\bar{\iota}^{\alpha'}
\end{pmatrix}.
\end{split}
\ee
One can simply package them as $k_a^A=(k^A_1,k_2^A)$ and $n_a^A=(n_1^A,n_2^A)$, where indices $a$ and $b$ are packaged from two basis spinors and are referred to as little group spinor indices. A rank-two tensor $\Omega_{AB}$ is introduced to raise or lower the spinor index \cite{Monteiro:2018xev}, which we decompose as 
\begin{equation}\label{Omega}
\Omega_{AB}=(k_{Aa}n_{Bb}-k_{Ba}n_{Bb})\epsilon^{ab}.
\end{equation}
One can verify that the right hand side of \eqref{Omega} fulfills all the properties of $\Omega_{AB}$.

The algebraic classification in the spinorial formalism has been constructed in \cite{Monteiro:2018xev}, which reproduces the full structure of the CMPP classification. For the five-dimensional type N spacetime, the non-zero little group four-spinors for the Weyl tensor are 
$\psi_{abcd}^{(4)}=C_{\mu \nu \rho \sigma} \sigma^{\mu \nu }_{\ \ AB}\sigma^{\rho \sigma}_{\ \ C D}n^A_{\ a}n^B_{\ b}n^C_{\ c}n^D_{\ d}$ \cite{Monteiro:2018xev}, where 
\be
\sigma^{\mu \nu }_{\ \ A B}=\frac{1}{2}(\gamma^\mu_{\ A C}\gamma^{\nu C}_{\ \ B}-\gamma^\nu_{\ A C}\gamma^{\mu C}_{\ \ B}).
\ee
A natural generalization of the WDC formula to five dimensions for the type N spacetime can be written as
\be
\psi_{abcd}^{(4)}\, \propto \,  \phi_{(ab}^{(2)}\phi_{cd)}^{(2)},
\ee
where $\phi_{ab}^{(2)}=F_{\mu \nu } \sigma^{\mu \nu }_{\ \ AB} n^A_{\ a}n^B_{\ b}$ are the non-zero little group two-spinors of a degenerate Maxwell tensor \cite{Monteiro:2018xev}. Here we have not specified the scalar field as the case in four dimensions since there are more non-zero Weyl scalars in five dimensions. In principle, there could be more scalar fields involved. Note that the four-dimensional case of the WDC relation for two equal electromagnetic fields is only valid for type D and type N spacetime where only one (complex) Weyl scalar is non-zero. More explicitly, we propose
\begin{equation}\label{WDC5d}
    \begin{split}
        &\psi_{1111}^{(4)}=\frac{3c}{S_1}\phi_{11}^{(2)}\phi_{11}^{(2)},\qquad \psi_{2222}^{(4)}=\frac{3c}{S_1'}\phi_{22}^{(2)}\phi_{22}^{(2)},\\
        &\psi_{1112}^{(4)}=\frac{3c}{S_2}\phi_{11}^{(2)}\phi_{12}^{(2)},\qquad \psi_{2221}^{(4)}=\frac{3c}{S_2'}\phi_{22}^{(2)}\phi_{21}^{(2)},\\ &\psi_{1212}^{(4)}=\frac{c}{S_3}[\phi_{11}^{(2)}\phi_{22}^{(2)}+2(\phi_{12}^{(2)})^2].
    \end{split}
\end{equation}
We have introduced two independent complex scalar fields $S_1$, $S_2$ and one real scalar field $S_3$. Then we will transform to a null frame to test the relations in \eqref{WDC5d} by the Bianchi identity and the Maxwell's equation.

In higher-dimensional spacetime, it is convenient to construct a null frame system for describing the classification of the Weyl tensor \cite{Coley:2004jv,Milson:2004jx,Pravda:2004ka,Ortaggio:2012jd}. The basis vectors are typically chosen as
\begin{equation}
\begin{split}
   & l=e_{0}=e^{1}, \quad n=e_{1}=e^{0},\\ & m_{i}=e_{i}=m^{i}=e^{i},\quad i=2,...,n-1,
\end{split}
\end{equation}
with two null vectors $l$ and $n$, and $n-2$ space-like vectors $m_{i}$. They satisfy the orthogonal conditions $l \cdot l =n \cdot n= l \cdot m^{i}= n\cdot m^{i}=0$, and the normalization conditions $l\cdot n=1,\, m^{i} \cdot m^{j}=\delta^{ij}$. The quantities $L_{\mu\nu}= \n_\nu l_\mu ,\, N_{\mu\nu}= \n_\nu n_\mu,\,  M_{\mu\nu}^k=\n_\nu  m^{k}_\mu$ are defined to specify the spin coefficients.

For a type N solution, the non-zero Weyl scalars are $\psi_{ij}=C_{\mu \nu \rho \sigma}n^\mu m^{i \nu} n^\rho m^{j \sigma}$ in the null frame \cite{Coley:2004jv,Milson:2004jx,Pravda:2004ka,Ortaggio:2012jd}, see also Appendix \ref{CMPP} and \ref{bispinor}. And $\phi_i=F_{\mu\nu}n^\mu m^{i\nu}$ are the non-zero Maxwell scalars of a degenerate Maxwell field. Converting the WDC formulas \eqref{WDC5d} to the null frame yields
\begin{equation}\label{WDC-type-N}
    \begin{split}
        &\psi_{22}-\psi_{33}-2i\psi_{23}=\frac{3c}{S_1}[(\phi_2)^2-(\phi_3)^2-2i \phi_2 \phi_3],\\
        &\psi_{22}-\psi_{33}+2i\psi_{23}=\frac{3c}{S_1'}[(\phi_2)^2-(\phi_3)^2 + 2i \phi_2 \phi_3],\\
        & \psi_{34} + i \psi_{24}=\frac{3c}{S_2}(\phi_3 \phi_4 + i \phi_2 \phi_4),\\
        & \psi_{34} - i \psi_{24}=\frac{3c}{S_2'}(\phi_3 \phi_4 - i \phi_2 \phi_4),\\
        &\psi_{44}=\frac{c}{S_3}[(\phi_2)^2+(\phi_3)^2+4(\phi_4)^2].
    \end{split}
\end{equation} 
The above equations present a natural extension of the WDC formulas to five dimensions for the type N spacetime. However, we are not able to extend the verification of \cite{Godazgar:2020zbv} to a generic five-dimensional type N spacetime. Because the equations for the scalar fields are highly entangled with the Maxwell scalars. We can not separate the equations for each scalar field. In the next section, we will show a self-contained reduction of the type N solution where one can realize a concrete WDC.

%%%%%%%%%%%%%%%%%%%%%%%%%%%%%%%%%%%%%%%%%%%%%%%%%%%%%%%%%%%%%%%%%%%%%%%%%%%%%%%%%%%%%%%%%%%%%%%%%%%%%%%%%%%%%%%%%%%%%%%%%%%%%%%%%%%%%%%%%%%%%%%%%%%%%%%%%%%%%%%%%%%%%%%%%%%%%%%%%%%%%%%%%%%%%%%%%%%%%%%%%%%%%%%%%%%%%%%%%%%%%%%%%%%%%%%%%%%%%%%%

\section{Reduced type N solution and the WDC relation}

We now consider a 4d-like reduction of the five-dimensional type N spacetime where we impose the restrictions $\psi_{4i}=0$ for the Weyl scalars.\footnote{Here, we refer to as a 4d-like reduction because we reduce the free Weyl scalars to the same amount as the four-dimensional case. This does not mean the extra dimension is trivial. Note also that the shear of the aligned null direction does not vanishes, which significantly distinguishes from the four-dimensional case.} The traceless property of the Weyl tensor then yields $\psi_{22}=-\psi_{33}$. Correspondingly, the Maxwell scalars and zeroth copies should be $\phi_4=0 , \,  S_3\rightarrow \infty$, where the divergent scalar field $S_3$ is chosen to turn off the extra component from the doubling in five dimensions. We postpone commenting on this curious point by the end of this section. A higher-dimensional Goldberg-Sachs-like theorem \cite{Durkee:2009nm,Ortaggio:2012hc} guarantees that for the CMPP type N spacetime the Weyl aligned null direction is geodesic (but no longer shear-free in general). Thus we can affine parameterize the geodesic and then take an appropriate spin transformation and a null rotation to make other basis vectors parallely propagated along $l$. We introduce the following definitions
\begin{equation}
\begin{split}
        &\Psi_4=-(\psi_{22}+i\psi_{23}), \quad \bar{\Psi}_4=-(\psi_{22}-i\psi_{23}), \\ &\varphi_2=-\frac{1}{\sqrt{2}}(\phi_2+i\phi_3),\quad \bar{\varphi}_2=-\frac{1}{\sqrt{2}}(\phi_2-i\phi_3),\\
        &\delta=\frac{1}{\sqrt{2}}(\delta_2-i\delta_3),\quad \bar{\delta}=\frac{1}{\sqrt{2}}(\delta_2+i\delta_3), \\ &\hat{\delta}=i\delta_4,\quad S=\frac{S_1}{c}.
\end{split}
\end{equation}
where $D$, $\Delta$, and $\delta_i$ are the directional derivatives associated to the basis vectors $l$, $n$, and $m_i$, respectively. Then, the Bianchi identity and the Maxwell's equation could be rewritten as (see in Appendix \ref{Maxwell} and \ref{Bianchi})
\begin{equation}
\begin{split}
 &D\Psi_4=-(L_{22}+iL_{23})\Psi_4,\\
&\delta\Psi_4=-\frac{1}{\sqrt{2}}\big[(2L_{12}+2M^2_{33}-L_{21}) \\
&\hspace{2.5cm}+i(L_{31}-2L_{13}+2M^2_{32})\big]\Psi_4,\\
&\hat{\delta}\Psi_4=[(M^3_{42}-2M^3_{24})+i(L_{41}-2L_{14}+M^2_{42})]\Psi_4,
\end{split}
\end{equation}
and
\begin{equation}
    \begin{split}
        &D\varphi_2=-(L_{22}+iL_{23})\varphi_2,\\
        &\delta\varphi_2=-\frac{1}{\sqrt{2}}\big[(L_{12}+M^2_{33}-L_{21})\\
        &\hspace{2.5cm}+i(L_{31}-L_{13}+M^2_{32})\big]\varphi_2,\\
        &\hat{\delta} \varphi_2=[(M^3_{42}-M^3_{24})+i(L_{41}-L_{14}+M^2_{42})]\varphi_2.
    \end{split}
\end{equation}
The WDC formula in the reduced case is simply given by
\begin{equation}\label{4d-like-WDC}
    \Psi_4=\frac{1}{S}(\varphi_2)^2.
\end{equation}
Now it is clear that the previously chosen divergent scalar field $S_3$ is to prevent the mixing term constructed from $\varphi_2 \bar\varphi_2$ in the Weyl doubling. Such a term is not involved at all in the four-dimensional WDC relation. Hence, it is reasonable to have a special treatment in five dimensions by simply imposing that there is no Weyl scalar associated to the $\varphi_2 \bar\varphi_2$ term.

Substituting the WDC relation \eqref{4d-like-WDC} into the Bianchi identity and simplifying them with the Maxwell's equation on the type N background, we obtain the equations for the scalar field
\begin{equation}\label{S}
\begin{split}
&D\log{S}=-(L_{22}+iL_{23}),\\   &\delta\log{S}=\frac{1}{\sqrt{2}}(L_{21}-iL_{31}),\\
&\hat{\delta}\log{S}=M^3_{42}+i(M^2_{42}+L_{41}).
\end{split}
\end{equation}
Following closely the treatment in \cite{Godazgar:2020zbv}, we have proven that the integrability conditions 
\begin{equation}
\begin{split}
    &(D\delta-\delta D)\log{S}=\frac{1}{\sqrt{2}}D(L_{21}-iL_{31})+\delta(L_{22}+iL_{23}),\\
    &(D\hat{\delta}-\hat{\delta}D)\log{S}=D(M^3_{42}+iM^2_{42}+iL_{41})+\hat{\delta}(L_{22}+iL_{23}),\\
    &(\delta\hat{\delta}-\hat{\delta}\delta)\log{S}=\delta(M^3_{42}+iM^2_{42}+iL_{41})-\frac{1}{\sqrt{2}}\hat{\delta}(L_{21}-iL_{31}).
\end{split}
\end{equation}
for the above differential equations are satisfied,
where the commutators of the operators $D$, $\delta_i$ in \cite{Coley:2004hu} and the Ricci identities in \cite{Ortaggio:2007eg} are applied. Finally, it is straightforward to verify that any solution of \eqref{S} solves the Klein-Gordon equation $\Box S=0$ on the type N background. This completes the zeroth copy of the five-dimensional WDC relation for the reduced type N solutions. In the next sections, we will present two examples of the five-dimensional WDC relation.

%%%%%%%%%%%%%%%%%%%%%%%%%%%%%%%%%%%%%%%%%%%%%%%%%%%%%%%%%%%%%%%%%%%%%%%%%%%%%%%%%%%%%%%%%%%%%%%%%%%%%%%%%%%%%%%%%%%%%%%%%%%%%%%%%%%%%%%%%%%%%%%%%%%%%%%%%%%%%%%%%%%%%%%%%%%%%%%%%%%%%%%%%%%%%%%%%%%%%%%%%%%%%%%%%%%%%%%%%%%%%%%%%%%%%%%%%%%%%%%%

\section{5d pp-wave solution} 

The line-element of the five-dimensional pp-wave solution with respect to the reduction introduced previously is given by \cite{Coley:2002ku}
\begin{align}
&\td s^2=2\td u [\td v + H(u,x_2,x_3,x_4) \td u] \nn\\
&\hspace{4cm} + \td x_2^{\ 2 } + \td x_3^{\ 2 }+ \td x_4^{\ 2 },\label{pp}\\
&\quad H(u,x_2,x_3,x_4)=f(u)x_4 + g(u,z)+\bar{g}(u,\bar{z}),\nn
\end{align}
where $z=x_2-ix_3$. We choose the same null frame system as \cite{Ortaggio:2023cdz}. A generic solution to \eqref{S} in the spacetime \eqref{pp} is $S=P(u,\bar{z})$. The Weyl scalar is given by $\Psi_4=\p_{\bar{z}}^2\bar{g}(u,\bar{z})$,
which allows us to determine the Maxwell scalar as
$\varphi_2=\sqrt{P(u,\bar{z})\p_{\bar{z}}^2\bar{g}(u,\bar{z})}$. We have checked directly that the Maxwell scalar satisfies the Maxwell's equation. Hence, a concrete WDC relation is constructed for the special pp-wave solution. It is easy to prove that the Maxwell field and the scalar field defined from the WDC for the pp-wave solution also satisfy the Maxwell's equation and the wave equation on five-dimensional Minkowski spacetime.

\section{5d Kundt solution} 

The type N Kundt solution in five dimensions with reduction is given by \cite{Coley:2006fr}
\begin{align}
&\td s^2=2 \td u [\td v + H(u,v,x^k) \td u + W_i(u,v,x^k) \td x^i] \nn \\
&\hspace{2cm}+ \delta_{ij} \td x^i \td x^j, \quad i,j,k=2,3,4, \label{kundt} \\
&\quad H(u,v,x^k)=\frac{v^2}{2y^2}+f(u,x^2,x^3)+g(u,x^2,x^4), \nn\\
&\quad  g(u,x^2,x^4)=\frac{1}{2}B_{43}(u)^2 (x^4)^2+g_1(u) x^2x^4 \nn \\
&\hspace{4cm} + B_{43}(u) C_3(u)x^4+g_2(u,x^2),\nn \\
&\quad  y=x^2,\quad W_2(u,v,x^k)=-2\frac{v}{y},\nn\\
&\quad  W_m(u,v,x^k)=x^nB_{nm}(u)+C_m(u),\quad m,n=3,4.\nn
\end{align}
Projecting the solution into the null frame system defined in \cite{Ortaggio:2023cdz}, one can obtain a generic solution to \eqref{S} for the Kundt solution \eqref{kundt} as $S= P(u,x_2-ix_3)/x_2$. The Weyl scalar of the Kundt solution is given by
\begin{multline}
        \Psi_4=\frac{1}{(x_2)^3}\bigg[(x_2+ix_3)B_{43}(u)^2-iB_{43}(u)C_4(u) -i\p_{x_3}f(u,x_2,x_3)
        \\
        -x_2\p_{x_3}^2f(u,x_2,x_3) +ix_2\p_{x_2}\p_{x_3}f(u,x_2,x_3)\bigg].
\end{multline}
Then, one can recover the Maxwell scalar from the square-root $\varphi_2=\sqrt{S\Psi_4}$. It can be verified directly that the Maxwell scalar satisfies the Maxwell’s equation. We further verify that the Maxwell and scalar fields defined from the Kundt solution also satisfy the Maxwell's equation and the wave equation on five-dimensional Minkowski spacetime. Note that the line-element of the Minkowski spacetime is now 
\be
\td s^2=2 \td u [\td v + \frac{v^2}{2y^2} \td u -2\frac{v}{y} \td y]  + \delta_{ij} \td x^i \td x^j.
\ee

%%%%%%%%%%%%%%%%%%%%%%%%%%%%%%%%%%%%%%%%%%%%%%%%%%%%%%%%%%%%%%%%%%%%%%%%%%%%%%%%%%%%%%%%%%%%%%%%%%%%%%%%%%%%%%%%%%%%%%%%%%%%%%%%%%%%%%%%%%%%%%%%%%%%%%%%%%%%%%%%%%%%%%%%%%%%%%%%%%%%%%%%%%%%%%%%%%%%%%%%%%%%%%%%%%%%%%%%%%%%%%%%%%%%%%%%%%%%%%%%

\section{Discussions} 

In this Letter, we offer the first realization of a higher-dimensional Weyl double copy relation, which significantly enlarges the scope of the WDC relation. The formulation of the WDC in four-dimensional spacetime is closely related to the algebraic classification. However, the algebraic classification of solutions in five dimensions is only relevant to the little group four-spinors of the Weyl tensor \cite{Monteiro:2018xev}. For type N spacetime, the little group four-spinor of the Weyl tensor is totally symmetric and is irreducible representations of $SU(2)$. Generically, the little group four-spinors of the Weyl tensor are decomposed into irreducible representations of $SU(2)$, which could consist of totally symmetry four-spinors, symmetric bi-spinors, and a scalar \cite{Monteiro:2018xev}. The WDC formula proposed in \eqref{WDC5d} is not enough for algebraically special spacetime of types other than N in five dimensions, nor even type N solutions in dimensions higher than five. One should also deal with the extra symmetric bi-spinors and scalar fields. Actually, this is the main obstacle for constructing the WDC in higher dimensions where the structure of the classification of the Weyl tensor is much richer than four dimensions. Nevertheless, the lesson from our construction is that one can always expect a 4d-like reduction where the WDC can be realized in any dimensions, at least, for algebraic type D and type N solutions, and the reduction would impose constraints on the components associated to all the extra dimensions. Such idea has been recently applied to recover the WDC for five-dimensional algebraic type D solutions \cite{Zhao:2024wtn}. At the end of the day, the reduction reveals the remarkable fact that interesting features in four-dimensional spacetime are compatible with extra dimensions. The reduction proposed for the WDC could provide a refined algebraic classification of higher-dimensional spacetime, which should be useful for finding new exact solutions in higher dimensions, see, e.g., the important application of the Kerr-Schild form in deriving higher dimensional rotating solutions \cite{Gibbons:2004uw,Gibbons:2004js}.

As a future direction, it is interesting to investigate the compatibility of the five-dimensional WDC with the Kerr-Schild double copy, which is vital to an important aspect of the classical double copy that the single and zeroth copies are solutions on the flat background spacetime or the full curved spacetime. The verification of the WDC relation in the present work (also the generic construction in \cite{Godazgar:2020zbv}) is purely at the equations of motion level in the vielbein formalism. We have never specified any exact solutions in a coordinates system. Then, the directional derivatives $D,\,\Delta,\,\delta_i$ are not specified and their connections to the directional derivatives associated to the flat background spacetime are not known. Hence, we can not use the WDC formulas to test any relations on the flat background spacetime. Nevertheless, we have verified for two exact solutions that the Maxwell fields and the scalar fields defined from the WDC formula also satisfy the Maxwell's equation and the wave equation on the flat background spacetime. Moreover, it is easy to verify that if one considers linearized gravity theory, our construction will lead to a WDC relation for the linearized Weyl tensor and the Maxwell tensor on the flat background spacetime, which is consistent with the investigations from the twistor perspective in \cite{White:2020sfn,Chacon:2021wbr}.

%%%%%%%%%%%%%%%%%%%%%%%%%%%%%%%%%%%%%%%%%%%%%%%%%%%%%%%%%%%%%%%%%%%%%%%%%%%%%%%%%%%%%%%%%%%%%%%%%%%%%%%%%%%%%%%%%%%%%%%%%%%%%%%%%%%%%%%%%%%%%%%%%%%%%%%%%%%%%%%%%%%%%%%%%%%%%%%%%%%%%%%%%%%%%%%%%%%%%%%%%%%%%%%%%%%%%%%%%%%%%%%%%%%%%%%%%%%%%%%%

\section*{Acknowledgments} 

The authors are grateful to Hong L\"{u} for useful discussions, to Zhengwen Liu, Ricardo Monteiro, and Chris White for very useful comments on the draft, and to Ricardo Monteiro again for useful correspondence on the integrability conditions. This work is supported in part by the National Natural Science Foundation of China (NSFC) under Grants No. 11935009, No. 12375066, and No. 12247103.

\appendix

\section{CMPP classification}
\label{CMPP}

Based on the null frame introduced in the main text, CMPP \cite{Coley:2004jv,Milson:2004jx,Pravda:2004ka,Ortaggio:2012jd} proposed an algebraic classification method for tensors.
Similar to the four-dimensional case, the CMPP classification is based on the maximum boost order of tensors. For any tensor $T_{\mu\nu...\rho}$, it can be projected onto the chosen null frame, and the boost weight of the null frame component $T_{ab...c}\equiv e^\mu_{\, {a}}e^\nu_{\, {b}}...e^\rho_{\, {c}}T_{\mu\nu...\rho}$ is defined as the number of the index $0$ minus the number of the index $1$ in the null frame component indices. The maximum boost weight of all non-zero null frame components is called the boost order of the tensor $T$. It could be proved that the boost order of a tensor depends only on the choice of null direction $l$. The essence of the CMPP classification is to choose a null vector $l$ to minimizes the maximum boost weight and another null vector $n$ that maximizes the minimum boost weight. 

In general, the Weyl tensor could be decomposed as
\begin{equation}
\begin{split}
&C_{\mu\nu\rho\sigma} = \ \underbrace{4C_{0i0j} n_{\{\mu} m_{\nu}^{i} n_{\rho} m_{\sigma\}}^{j}}_{\text{boost weight } +2}
+ \underbrace{8C_{010i} n_{\{\mu} l_{\nu} n_{\rho} m_{\sigma\}}^{i}+4C_{0ijk} n_{\{\mu} m_{\nu}^{i} m_{\rho}^{j} m_{\sigma\}}^{k}}_{+1}\\
&+ \underbrace{4C_{0101} n_{\{\mu} l_{\nu} n_{\rho} l_{\sigma\}}
+4C_{01ij} n_{\{\mu} l_{\nu} m_{\rho}^{i} m_{\sigma\}}^{j}
+8C_{0i1j} n_{\{\mu} m_{\nu}^{i} l_{\rho} m_{\sigma\}}^{j}
+ C_{ijkl} m_{\{\mu}^{i} m_{\nu}^{j} m_{\rho}^{k} m_{\sigma\}}^{l}}_{0}\\
&\hspace{2cm} + \underbrace{8C_{101i} l_{\{\mu} n_{\nu} l_{\rho} m_{\sigma\}}^{i}
+ 4C_{1ijk} l_{\{\mu} m_{\nu}^{i} m_{\rho}^{j} m_{\sigma\}}^{k}}_{-1}
+ \underbrace{4C_{1i1j} l_{\{\mu} m_{\nu}^{i} l_{\rho} m_{\sigma\}}^{j}}_{-2}.
\end{split}
\end{equation}
where the notation $C_{\mu\nu\rho\sigma}=C_{\{\mu\nu\rho\sigma\}}\equiv\frac{1}{2}(C_{[\mu\nu][\rho\sigma]}+C_{[\rho\sigma][\mu\nu]})
$ should be understood. Specifically, a type N spacetime in the CMPP classification is defined when its Weyl tensor has the following form
\begin{equation}
C_{\mu\nu\rho\sigma}=4C_{1i1j} l_{\{\mu} m_{\nu}^{i} l_{\rho} m_{\sigma\}}^{j}\equiv 8\psi_{ij}l_{\{\mu} m_{\nu}^{i} l_{\rho} m_{\sigma\}}^{j}.
\end{equation}
In the null frame, the Maxwell tensor can be decomposed as
\begin{equation}
        F_{\mu \nu}=F_{0i}(n_{\mu}m^i_\nu-n_\nu m^i_\mu)+F_{ij}m^i_\mu m^j_\nu+F_{01}(n_\mu l_\nu-l_\mu n_\mu)+F_{1i}(l_\mu m^i_\nu-l_\nu m^i_\mu),
\end{equation}
where the boost weights of the right hand side terms are 1, 0, 0, -1, respectively. A degenerate (type N) Maxwell tensor in the CMPP classification should have the following form 
\begin{equation}
        F_{\mu \nu}=F_{1i}(l_\mu m^i_\nu-l_\nu m^i_\mu)\equiv \phi_i(l_\mu m^i_\nu-l_\nu m^i_\mu).
\end{equation}

\section{Connections between the little group bi-spinors and the spacetime tensors}
\label{bispinor}

To convert from the spinorial space to the null frame, we need to apply the following relation 
\begin{equation}
\begin{split}
    \gamma^\mu_{\ AC}\gamma^{\nu C}_{\ \ B} n^A_{\ a}n^B_{\ b}&=-\gamma^\mu_{\ AD}n^A_{\ a}n^{Dd}\gamma^\nu_{EB}k^E_{\ d}n^B_{\ b}+\gamma^\mu_{\ AD}n^A_{\ a}k^D_{\ d}\gamma^\nu_{\ EB}n^{Ed}n^B_{\ b}\\
     &=-\sqrt{2}n^\mu \delta_a^{\ d} \varepsilon^\nu_{\ db}-\sqrt{2}\varepsilon^\mu_{\ ad}n^\nu\delta^D_{\ b} \\&=\sqrt{2}\varepsilon^{[\mu}_{\ ab}n^{\nu]},
     \end{split}
\end{equation}
where $\varepsilon^{\mu}_{\ ab}=k_a^A \gamma^\mu_{\ AC}n_b^C$ \cite{Monteiro:2018xev}. In particular,
\begin{equation}
    \varepsilon^\mu_{11}=\frac{1}{\sqrt{2}}(m^{2\mu}-im^{3\mu}),\quad \varepsilon^\mu_{22}=\frac{1}{\sqrt{2}}(m^{2\mu}+im^{3\mu}), \quad \varepsilon^\mu_{12}=\varepsilon^\mu_{21}=im^{4\mu},
\end{equation}
where $m^{2\mu},m^{3\mu},m^{4\mu}$ are three spacelike vectors in the null frame. Finally, we uncover the connections between the little group bi-spinors and the spacetime tensors as
\begin{equation}
        \psi_{abcd}^{(4)}=C_{\mu \nu \rho \sigma}\sigma^{\mu \nu}_{\ \ AB}\sigma^{\rho \sigma}_{\ \ CD}n^A_{\ a}n^B_{\ b}n^C_{\ c}n^D_{\ d} 
            =2C_{[\mu \nu] [\rho \sigma]}\varepsilon^\mu_{\ ab}n^\nu \varepsilon^\rho_{\ cd}n^\sigma,
\end{equation}
and
\be
\phi_{ab}^{(2)}=F_{\mu \nu}\varepsilon^\mu_{\ ab}n^\nu .
\ee

\section{Maxwell's equation in the null frame}
\label{Maxwell}

\begin{equation}
    DF_{01}+\delta_iF_{0i}+F_{0i}(M^i_{jj}-L_{1i}-N_{i0})+F_{01}L_{ii}+F_{ij}L_{ji}+F_{1i}L_{i0}=0,
\end{equation}
\begin{equation}
    \Delta F_{01}-\delta_iF_{1i}+F_{01}N_{ii}-F_{ij}N_{ji}-F_{0i}N_{i1}+F_{1i}(L_{i1}-L_{1i}-M^i_{jj})=0,
\end{equation}
\begin{multline}
    \Delta F_{0i}+\delta_k F_{ki}+DF_{1i}+F_{0i}(N_{jj}-L_{11})+F_{1i}(L_{10}+L_{jj})+F_{ki}(M^i_{jj}-N_{k0}-L_{k1})\\+F_{0k}(M^k_{i1}-N_{ik})+F_{kj}M^j_{ik}+F_{01}(L_{i1}-N_{i0})+F_{1k}(M^k_{i0}-L_{ik})=0,
\end{multline}
\begin{multline}
    DF_{1i}-\Delta F_{0i}+\delta_iF_{01}=-F_{0i}L_{11}+F_{0j}(N_{ji}-M^i_{j1})+F_{ij}(L_{j1}-N_{j0})\\ +F_{01}(N_{i0}+L_{i1})-F_{1i}L_{10}+F_{1j}(M^i_{j0}-L_{ji}),
\end{multline}
\begin{multline}
    DF_{ij}-2\delta_{[i|}F_{0|j]}=2F_{0[i|}L_{1|j]}-2F_{0k}M^k_{[ij]}+2F_{0[i}N_{j]0}-2F_{01}L_{[ij]} \\ -2F_{[i|k}L_{k|j]}-2F_{[i|k|}M^k_{j]0}+2F_{1[i}L_{j]0},
\end{multline}
\begin{multline}
    \Delta F_{ij}-2\delta_{[i|}F_{1|j]}=2F_{0[i}N_{j]1}+2F_{01}N_{[ij]}+2F_{k[i}M^k_{j]1} \\ +2F_{k[i|}N_{k|j]}+2F_{1[i}L_{j]1}-2F_{1[i|}L_{1|j]}-2F_{1k}M^k_{[ij]},
\end{multline}
\begin{equation}
    \delta_iF_{jk}=2F_{1i}L_{jk}+2F_{0i}N_{jk}+2F_{li}M^l_{jk},
\end{equation}
where the permutation of the indices $i,j,k$ for the last equation should be understood.

\section{Bianchi identity and Maxwell's equation with reduction}
\label{Bianchi}

The Bianchi identity in the null frame \cite{Pravda:2004ka} are now reduced to
\begin{equation}
    \begin{split}
        &D\psi_{ij}=-\psi_{ik}L_{kj},\\
        &\delta_{[k}\psi_{j]i}=-\psi_{il}M^l_{[jk]}+M^l_{i[j}\psi_{k]l}+\psi_{i[j}L_{k]1}+2L_{1[j}\psi_{k]i},\\
        &L_{k[i}\psi_{j]k}=0,\quad L_{k[j}\psi_{m]i}+L_{i[m}\psi_{j]k}=0,\quad \psi_{i[k}L_{jm]}=0,
    \end{split}
\end{equation}
which yield 
\begin{equation}
    \begin{split}
    &L_{22}=L_{33},\quad L_{32}=-L_{23},\quad L_{i4}=L_{4i}=0,\\
    &M^2_{44}=M^3_{44}=0,\quad M^2_{42}=M^3_{43},\quad M^2_{43}=-M^3_{42}.
    \end{split}
\end{equation}
With the above reduction, the Maxwell's equation is now reduced to
\begin{equation}
\begin{split}
    &D \phi_{i}=-\phi_jL_{ji},\\
    &\delta_j\phi_i-\delta_i\phi_j=-\phi_i L_{1j}+\phi_j L_{1i}-\phi_j L_{i1}+\phi_i L_{j1}+\phi_k M^k_{ji}-\phi_kM^k_{ij},\\
    &\delta_i\phi_i=-\phi_iL_{1i}+\phi_iL_{i1}\phi_iM^i_{jj}.
\end{split}
\end{equation}

\providecommand{\href}[2]{#2}\begingroup\raggedright\endgroup

%\bibliography{ref}

\begin{thebibliography}{10}

\bibitem{Maldacena:1997re}
J.~M. Maldacena, ``{The Large N limit of superconformal field theories and
  supergravity},'' \href{http://dx.doi.org/10.4310/ATMP.1998.v2.n2.a1}{{\em
  Adv. Theor. Math. Phys.} {\bfseries 2} (1998) 231--252},
  \href{http://arxiv.org/abs/hep-th/9711200}{{\ttfamily arXiv:hep-th/9711200}}.

\bibitem{Kawai:1985xq}
H.~Kawai, D.~C. Lewellen, and S.~H.~H. Tye, ``{A Relation Between Tree
  Amplitudes of Closed and Open Strings},''
  \href{http://dx.doi.org/10.1016/0550-3213(86)90362-7}{{\em Nucl. Phys. B}
  {\bfseries 269} (1986) 1--23}.

\bibitem{Bern:2008qj}
Z.~Bern, J.~J.~M. Carrasco, and H.~Johansson, ``{New Relations for Gauge-Theory
  Amplitudes},'' \href{http://dx.doi.org/10.1103/PhysRevD.78.085011}{{\em Phys.
  Rev. D} {\bfseries 78} (2008) 085011},
  \href{http://arxiv.org/abs/0805.3993}{{\ttfamily arXiv:0805.3993 [hep-ph]}}.

\bibitem{Bern:2010ue}
Z.~Bern, J.~J.~M. Carrasco, and H.~Johansson, ``{Perturbative Quantum Gravity
  as a Double Copy of Gauge Theory},''
  \href{http://dx.doi.org/10.1103/PhysRevLett.105.061602}{{\em Phys. Rev.
  Lett.} {\bfseries 105} (2010) 061602},
  \href{http://arxiv.org/abs/1004.0476}{{\ttfamily arXiv:1004.0476 [hep-th]}}.

\bibitem{Bern:2019prr}
Z.~Bern, J.~J. Carrasco, M.~Chiodaroli, H.~Johansson, and R.~Roiban, ``{The
  duality between color and kinematics and its applications},''
  \href{http://dx.doi.org/10.1088/1751-8121/ad5fd0}{{\em J. Phys. A} {\bfseries
  57} no.~33, (2024) 333002}, \href{http://arxiv.org/abs/1909.01358}{{\ttfamily
  arXiv:1909.01358 [hep-th]}}.

\bibitem{Bern:2022wqg}
Z.~Bern, J.~J. Carrasco, M.~Chiodaroli, H.~Johansson, and R.~Roiban, ``{The
  SAGEX review on scattering amplitudes Chapter 2: An invitation to
  color-kinematics duality and the double copy},''
  \href{http://dx.doi.org/10.1088/1751-8121/ac93cf}{{\em J. Phys. A} {\bfseries
  55} no.~44, (2022) 443003}, \href{http://arxiv.org/abs/2203.13013}{{\ttfamily
  arXiv:2203.13013 [hep-th]}}.

\bibitem{White:2017mwc}
C.~D. White, ``{The double copy: gravity from gluons},''
  \href{http://dx.doi.org/10.1080/00107514.2017.1415725}{{\em Contemp. Phys.}
  {\bfseries 59} (2018) 109}, \href{http://arxiv.org/abs/1708.07056}{{\ttfamily
  arXiv:1708.07056 [hep-th]}}.

\bibitem{Adamo:2022dcm}
T.~Adamo, J.~J.~M. Carrasco, M.~Carrillo-Gonz\'alez, M.~Chiodaroli, H.~Elvang,
  H.~Johansson, D.~O'Connell, R.~Roiban, and O.~Schlotterer, ``{Snowmass White
  Paper: the Double Copy and its Applications},'' in {\em {Snowmass 2021}}.
\newblock 4, 2022.
\newblock \href{http://arxiv.org/abs/2204.06547}{{\ttfamily arXiv:2204.06547
  [hep-th]}}.

\bibitem{Bern:2019nnu}
Z.~Bern, C.~Cheung, R.~Roiban, C.-H. Shen, M.~P. Solon, and M.~Zeng,
  ``{Scattering Amplitudes and the Conservative Hamiltonian for Binary Systems
  at Third Post-Minkowskian Order},''
  \href{http://dx.doi.org/10.1103/PhysRevLett.122.201603}{{\em Phys. Rev.
  Lett.} {\bfseries 122} no.~20, (2019) 201603},
  \href{http://arxiv.org/abs/1901.04424}{{\ttfamily arXiv:1901.04424
  [hep-th]}}.

\bibitem{Bern:2019crd}
Z.~Bern, C.~Cheung, R.~Roiban, C.-H. Shen, M.~P. Solon, and M.~Zeng, ``{Black
  Hole Binary Dynamics from the Double Copy and Effective Theory},''
  \href{http://dx.doi.org/10.1007/JHEP10(2019)206}{{\em JHEP} {\bfseries 10}
  (2019) 206}, \href{http://arxiv.org/abs/1908.01493}{{\ttfamily
  arXiv:1908.01493 [hep-th]}}.

\bibitem{Bern:2021dqo}
Z.~Bern, J.~Parra-Martinez, R.~Roiban, M.~S. Ruf, C.-H. Shen, M.~P. Solon, and
  M.~Zeng, ``{Scattering Amplitudes and Conservative Binary Dynamics at ${\cal
  O}(G^4)$},'' \href{http://dx.doi.org/10.1103/PhysRevLett.126.171601}{{\em
  Phys. Rev. Lett.} {\bfseries 126} no.~17, (2021) 171601},
  \href{http://arxiv.org/abs/2101.07254}{{\ttfamily arXiv:2101.07254
  [hep-th]}}.

\bibitem{Bern:2021yeh}
Z.~Bern, J.~Parra-Martinez, R.~Roiban, M.~S. Ruf, C.-H. Shen, M.~P. Solon, and
  M.~Zeng, ``{Scattering Amplitudes, the Tail Effect, and Conservative Binary
  Dynamics at O(G4)},''
  \href{http://dx.doi.org/10.1103/PhysRevLett.128.161103}{{\em Phys. Rev.
  Lett.} {\bfseries 128} no.~16, (2022) 161103},
  \href{http://arxiv.org/abs/2112.10750}{{\ttfamily arXiv:2112.10750
  [hep-th]}}.

\bibitem{Monteiro:2014cda}
R.~Monteiro, D.~O'Connell, and C.~D. White, ``{Black holes and the double
  copy},'' \href{http://dx.doi.org/10.1007/JHEP12(2014)056}{{\em JHEP}
  {\bfseries 12} (2014) 056}, \href{http://arxiv.org/abs/1410.0239}{{\ttfamily
  arXiv:1410.0239 [hep-th]}}.

\bibitem{Luna:2015paa}
A.~Luna, R.~Monteiro, D.~O'Connell, and C.~D. White, ``{The classical double
  copy for Taub\textendash{}NUT spacetime},''
  \href{http://dx.doi.org/10.1016/j.physletb.2015.09.021}{{\em Phys. Lett. B}
  {\bfseries 750} (2015) 272--277},
  \href{http://arxiv.org/abs/1507.01869}{{\ttfamily arXiv:1507.01869
  [hep-th]}}.

\bibitem{Luna:2016due}
A.~Luna, R.~Monteiro, I.~Nicholson, D.~O'Connell, and C.~D. White, ``{The
  double copy: Bremsstrahlung and accelerating black holes},''
  \href{http://dx.doi.org/10.1007/JHEP06(2016)023}{{\em JHEP} {\bfseries 06}
  (2016) 023}, \href{http://arxiv.org/abs/1603.05737}{{\ttfamily
  arXiv:1603.05737 [hep-th]}}.

\bibitem{Goldberger:2016iau}
W.~D. Goldberger and A.~K. Ridgway, ``{Radiation and the classical double copy
  for color charges},''
  \href{http://dx.doi.org/10.1103/PhysRevD.95.125010}{{\em Phys. Rev. D}
  {\bfseries 95} no.~12, (2017) 125010},
  \href{http://arxiv.org/abs/1611.03493}{{\ttfamily arXiv:1611.03493
  [hep-th]}}.

\bibitem{Goldberger:2017frp}
W.~D. Goldberger, S.~G. Prabhu, and J.~O. Thompson, ``{Classical gluon and
  graviton radiation from the bi-adjoint scalar double copy},''
  \href{http://dx.doi.org/10.1103/PhysRevD.96.065009}{{\em Phys. Rev. D}
  {\bfseries 96} no.~6, (2017) 065009},
  \href{http://arxiv.org/abs/1705.09263}{{\ttfamily arXiv:1705.09263
  [hep-th]}}.

\bibitem{Bahjat-Abbas:2017htu}
N.~Bahjat-Abbas, A.~Luna, and C.~D. White, ``{The Kerr-Schild double copy in
  curved spacetime},'' \href{http://dx.doi.org/10.1007/JHEP12(2017)004}{{\em
  JHEP} {\bfseries 12} (2017) 004},
  \href{http://arxiv.org/abs/1710.01953}{{\ttfamily arXiv:1710.01953
  [hep-th]}}.

\bibitem{Carrillo-Gonzalez:2017iyj}
M.~Carrillo-Gonz\'alez, R.~Penco, and M.~Trodden, ``{The classical double copy
  in maximally symmetric spacetimes},''
  \href{http://dx.doi.org/10.1007/JHEP04(2018)028}{{\em JHEP} {\bfseries 04}
  (2018) 028}, \href{http://arxiv.org/abs/1711.01296}{{\ttfamily
  arXiv:1711.01296 [hep-th]}}.

\bibitem{Shen:2018ebu}
C.-H. Shen, ``{Gravitational Radiation from Color-Kinematics Duality},''
  \href{http://dx.doi.org/10.1007/JHEP11(2018)162}{{\em JHEP} {\bfseries 11}
  (2018) 162}, \href{http://arxiv.org/abs/1806.07388}{{\ttfamily
  arXiv:1806.07388 [hep-th]}}.

\bibitem{Lee:2018gxc}
K.~Lee, ``{Kerr-Schild Double Field Theory and Classical Double Copy},''
  \href{http://dx.doi.org/10.1007/JHEP10(2018)027}{{\em JHEP} {\bfseries 10}
  (2018) 027}, \href{http://arxiv.org/abs/1807.08443}{{\ttfamily
  arXiv:1807.08443 [hep-th]}}.

\bibitem{Berman:2018hwd}
D.~S. Berman, E.~Chac\'on, A.~Luna, and C.~D. White, ``{The self-dual classical
  double copy, and the Eguchi-Hanson instanton},''
  \href{http://dx.doi.org/10.1007/JHEP01(2019)107}{{\em JHEP} {\bfseries 01}
  (2019) 107}, \href{http://arxiv.org/abs/1809.04063}{{\ttfamily
  arXiv:1809.04063 [hep-th]}}.

\bibitem{CarrilloGonzalez:2019gof}
M.~Carrillo~Gonz\'alez, B.~Melcher, K.~Ratliff, S.~Watson, and C.~D. White,
  ``{The classical double copy in three spacetime dimensions},''
  \href{http://dx.doi.org/10.1007/JHEP07(2019)167}{{\em JHEP} {\bfseries 07}
  (2019) 167}, \href{http://arxiv.org/abs/1904.11001}{{\ttfamily
  arXiv:1904.11001 [hep-th]}}.

\bibitem{Easson:2020esh}
D.~A. Easson, C.~Keeler, and T.~Manton, ``{Classical double copy of nonsingular
  black holes},'' \href{http://dx.doi.org/10.1103/PhysRevD.102.086015}{{\em
  Phys. Rev. D} {\bfseries 102} no.~8, (2020) 086015},
  \href{http://arxiv.org/abs/2007.16186}{{\ttfamily arXiv:2007.16186 [gr-qc]}}.

\bibitem{Gumus:2020hbb}
M.~K. Gumus and G.~Alkac, ``{More on the classical double copy in three
  spacetime dimensions},''
  \href{http://dx.doi.org/10.1103/PhysRevD.102.024074}{{\em Phys. Rev. D}
  {\bfseries 102} no.~2, (2020) 024074},
  \href{http://arxiv.org/abs/2006.00552}{{\ttfamily arXiv:2006.00552
  [hep-th]}}.

\bibitem{Almeida:2020mrg}
G.~L. Almeida, S.~Foffa, and R.~Sturani, ``{Classical Gravitational Self-Energy
  from Double Copy},'' \href{http://dx.doi.org/10.1007/JHEP11(2020)165}{{\em
  JHEP} {\bfseries 11} (2020) 165},
  \href{http://arxiv.org/abs/2008.06195}{{\ttfamily arXiv:2008.06195 [gr-qc]}}.

\bibitem{Campiglia:2021srh}
M.~Campiglia and S.~Nagy, ``{A double copy for asymptotic symmetries in the
  self-dual sector},'' \href{http://dx.doi.org/10.1007/JHEP03(2021)262}{{\em
  JHEP} {\bfseries 03} (2021) 262},
  \href{http://arxiv.org/abs/2102.01680}{{\ttfamily arXiv:2102.01680
  [hep-th]}}.

\bibitem{Alkac:2021bav}
G.~Alkac, M.~K. Gumus, and M.~Tek, ``{The Kerr-Schild Double Copy in Lifshitz
  Spacetime},'' \href{http://dx.doi.org/10.1007/JHEP05(2021)214}{{\em JHEP}
  {\bfseries 05} (2021) 214}, \href{http://arxiv.org/abs/2103.06986}{{\ttfamily
  arXiv:2103.06986 [hep-th]}}.

\bibitem{Alkac:2021seh}
G.~Alkac, M.~K. Gumus, and M.~A. Olpak, ``{Kerr-Schild double copy of the
  Coulomb solution in three dimensions},''
  \href{http://dx.doi.org/10.1103/PhysRevD.104.044034}{{\em Phys. Rev. D}
  {\bfseries 104} no.~4, (2021) 044034},
  \href{http://arxiv.org/abs/2105.11550}{{\ttfamily arXiv:2105.11550
  [hep-th]}}.

\bibitem{Chacon:2021hfe}
E.~Chac\'on, A.~Luna, and C.~D. White, ``{Double copy of the multipole
  expansion},'' \href{http://dx.doi.org/10.1103/PhysRevD.106.086020}{{\em Phys.
  Rev. D} {\bfseries 106} no.~8, (2022) 086020},
  \href{http://arxiv.org/abs/2108.07702}{{\ttfamily arXiv:2108.07702
  [hep-th]}}.

\bibitem{Adamo:2021dfg}
T.~Adamo and U.~Kol, ``{Classical double copy at null infinity},''
  \href{http://dx.doi.org/10.1088/1361-6382/ac635e}{{\em Class. Quant. Grav.}
  {\bfseries 39} no.~10, (2022) 105007},
  \href{http://arxiv.org/abs/2109.07832}{{\ttfamily arXiv:2109.07832
  [hep-th]}}.

\bibitem{Mao:2021kxq}
P.~Mao and W.~Zhao, ``{Note on the asymptotic structure of Kerr-Schild form},''
  \href{http://dx.doi.org/10.1007/JHEP01(2022)030}{{\em JHEP} {\bfseries 01}
  (2022) 030}, \href{http://arxiv.org/abs/2109.09676}{{\ttfamily
  arXiv:2109.09676 [gr-qc]}}.

\bibitem{Shi:2021qsb}
C.~Shi and J.~Plefka, ``{Classical double copy of worldline quantum field
  theory},'' \href{http://dx.doi.org/10.1103/PhysRevD.105.026007}{{\em Phys.
  Rev. D} {\bfseries 105} no.~2, (2022) 026007},
  \href{http://arxiv.org/abs/2109.10345}{{\ttfamily arXiv:2109.10345
  [hep-th]}}.

\bibitem{Didenko:2022qxq}
V.~E. Didenko and N.~K. Dosmanbetov, ``{Classical Double Copy and Higher-Spin
  Fields},'' \href{http://dx.doi.org/10.1103/PhysRevLett.130.071603}{{\em Phys.
  Rev. Lett.} {\bfseries 130} no.~7, (2023) 071603},
  \href{http://arxiv.org/abs/2210.04704}{{\ttfamily arXiv:2210.04704
  [hep-th]}}.

\bibitem{Chawla:2023bsu}
S.~Chawla and C.~Keeler, ``{Black hole horizons from the double copy},''
  \href{http://dx.doi.org/10.1088/1361-6382/acfe57}{{\em Class. Quant. Grav.}
  {\bfseries 40} no.~22, (2023) 225004},
  \href{http://arxiv.org/abs/2306.02417}{{\ttfamily arXiv:2306.02417
  [hep-th]}}.

\bibitem{Ceresole:2023wxg}
A.~Ceresole, T.~Damour, A.~Nagar, and P.~Rettegno, ``{Double copy, Kerr-Schild
  gauges and the Effective-One-Body formalism},''
  \href{http://arxiv.org/abs/2312.01478}{{\ttfamily arXiv:2312.01478 [gr-qc]}}.

\bibitem{Ferrero:2024eva}
P.~Ferrero, D.~Francia, C.~Heissenberg, and M.~Romoli, ``{Double-copy
  supertranslations},''
  \href{http://dx.doi.org/10.1103/PhysRevD.110.026009}{{\em Phys. Rev. D}
  {\bfseries 110} no.~2, (2024) 026009},
  \href{http://arxiv.org/abs/2402.11595}{{\ttfamily arXiv:2402.11595
  [hep-th]}}.

\bibitem{Luna:2018dpt}
A.~Luna, R.~Monteiro, I.~Nicholson, and D.~O'Connell, ``{Type D Spacetimes and
  the Weyl Double Copy},''
  \href{http://dx.doi.org/10.1088/1361-6382/ab03e6}{{\em Class. Quant. Grav.}
  {\bfseries 36} (2019) 065003},
  \href{http://arxiv.org/abs/1810.08183}{{\ttfamily arXiv:1810.08183
  [hep-th]}}.

\bibitem{Keeler:2020rcv}
C.~Keeler, T.~Manton, and N.~Monga, ``{From Navier-Stokes to Maxwell via
  Einstein},'' \href{http://dx.doi.org/10.1007/JHEP08(2020)147}{{\em JHEP}
  {\bfseries 08} (2020) 147}, \href{http://arxiv.org/abs/2005.04242}{{\ttfamily
  arXiv:2005.04242 [hep-th]}}.

\bibitem{Alawadhi:2020jrv}
R.~Alawadhi, D.~S. Berman, and B.~Spence, ``{Weyl doubling},''
  \href{http://dx.doi.org/10.1007/JHEP09(2020)127}{{\em JHEP} {\bfseries 09}
  (2020) 127}, \href{http://arxiv.org/abs/2007.03264}{{\ttfamily
  arXiv:2007.03264 [hep-th]}}.

\bibitem{Godazgar:2020zbv}
H.~Godazgar, M.~Godazgar, R.~Monteiro, D.~Peinador~Veiga, and C.~N. Pope,
  ``{Weyl Double Copy for Gravitational Waves},''
  \href{http://dx.doi.org/10.1103/PhysRevLett.126.101103}{{\em Phys. Rev.
  Lett.} {\bfseries 126} no.~10, (2021) 101103},
  \href{http://arxiv.org/abs/2010.02925}{{\ttfamily arXiv:2010.02925
  [hep-th]}}.

\bibitem{White:2020sfn}
C.~D. White, ``{Twistorial Foundation for the Classical Double Copy},''
  \href{http://dx.doi.org/10.1103/PhysRevLett.126.061602}{{\em Phys. Rev.
  Lett.} {\bfseries 126} no.~6, (2021) 061602},
  \href{http://arxiv.org/abs/2012.02479}{{\ttfamily arXiv:2012.02479
  [hep-th]}}.

\bibitem{Monteiro:2020plf}
R.~Monteiro, D.~O'Connell, D.~Peinador~Veiga, and M.~Sergola, ``{Classical
  solutions and their double copy in split signature},''
  \href{http://dx.doi.org/10.1007/JHEP05(2021)268}{{\em JHEP} {\bfseries 05}
  (2021) 268}, \href{http://arxiv.org/abs/2012.11190}{{\ttfamily
  arXiv:2012.11190 [hep-th]}}.

\bibitem{Chacon:2021wbr}
E.~Chac\'on, S.~Nagy, and C.~D. White, ``{The Weyl double copy from twistor
  space},'' \href{http://dx.doi.org/10.1007/JHEP05(2021)239}{{\em JHEP}
  {\bfseries 05} (2021) 239}, \href{http://arxiv.org/abs/2103.16441}{{\ttfamily
  arXiv:2103.16441 [hep-th]}}.

\bibitem{Godazgar:2021iae}
H.~Godazgar, M.~Godazgar, R.~Monteiro, D.~Peinador~Veiga, and C.~N. Pope,
  ``{Asymptotic Weyl double copy},''
  \href{http://dx.doi.org/10.1007/JHEP11(2021)126}{{\em JHEP} {\bfseries 11}
  (2021) 126}, \href{http://arxiv.org/abs/2109.07866}{{\ttfamily
  arXiv:2109.07866 [hep-th]}}.

\bibitem{Easson:2021asd}
D.~A. Easson, T.~Manton, and A.~Svesko, ``{Sources in the Weyl Double Copy},''
  \href{http://dx.doi.org/10.1103/PhysRevLett.127.271101}{{\em Phys. Rev.
  Lett.} {\bfseries 127} no.~27, (2021) 271101},
  \href{http://arxiv.org/abs/2110.02293}{{\ttfamily arXiv:2110.02293 [gr-qc]}}.

\bibitem{Chacon:2021lox}
E.~Chac\'on, S.~Nagy, and C.~D. White, ``{Alternative formulations of the
  twistor double copy},'' \href{http://dx.doi.org/10.1007/JHEP03(2022)180}{{\em
  JHEP} {\bfseries 03} (2022) 180},
  \href{http://arxiv.org/abs/2112.06764}{{\ttfamily arXiv:2112.06764
  [hep-th]}}.

\bibitem{Han:2022ubu}
S.~Han, ``{Weyl double copy and massless free-fields in curved spacetimes},''
  \href{http://dx.doi.org/10.1088/1361-6382/ac96c2}{{\em Class. Quant. Grav.}
  {\bfseries 39} no.~22, (2022) 225009},
  \href{http://arxiv.org/abs/2204.01907}{{\ttfamily arXiv:2204.01907 [gr-qc]}}.

\bibitem{Han:2022mze}
S.~Han, ``{The Weyl double copy in vacuum spacetimes with a cosmological
  constant},'' \href{http://dx.doi.org/10.1007/JHEP09(2022)238}{{\em JHEP}
  {\bfseries 09} (2022) 238}, \href{http://arxiv.org/abs/2205.08654}{{\ttfamily
  arXiv:2205.08654 [gr-qc]}}.

\bibitem{Luna:2022dxo}
A.~Luna, N.~Moynihan, and C.~D. White, ``{Why is the Weyl double copy local in
  position space?},'' \href{http://dx.doi.org/10.1007/JHEP12(2022)046}{{\em
  JHEP} {\bfseries 12} (2022) 046},
  \href{http://arxiv.org/abs/2208.08548}{{\ttfamily arXiv:2208.08548
  [hep-th]}}.

\bibitem{Easson:2022zoh}
D.~A. Easson, T.~Manton, and A.~Svesko, ``{Einstein-Maxwell theory and the Weyl
  double copy},'' \href{http://dx.doi.org/10.1103/PhysRevD.107.044063}{{\em
  Phys. Rev. D} {\bfseries 107} no.~4, (2023) 044063},
  \href{http://arxiv.org/abs/2210.16339}{{\ttfamily arXiv:2210.16339 [gr-qc]}}.

\bibitem{Easson:2023dbk}
D.~A. Easson, G.~Herczeg, T.~Manton, and M.~Pezzelle, ``{Isometries and the
  double copy},'' \href{http://dx.doi.org/10.1007/JHEP09(2023)162}{{\em JHEP}
  {\bfseries 09} (2023) 162}, \href{http://arxiv.org/abs/2306.13687}{{\ttfamily
  arXiv:2306.13687 [gr-qc]}}.

\bibitem{Alkac:2023glx}
G.~Alkac, M.~K. Gumus, O.~Kasikci, M.~A. Olpak, and M.~Tek, ``{Regularized Weyl
  double copy},'' \href{http://dx.doi.org/10.1103/PhysRevD.109.084047}{{\em
  Phys. Rev. D} {\bfseries 109} no.~8, (2024) 084047},
  \href{http://arxiv.org/abs/2310.06048}{{\ttfamily arXiv:2310.06048
  [hep-th]}}.

\bibitem{Mao:2023yle}
P.~Mao and W.~Zhao, ``{Asymptotic Weyl double copy in Newman-Penrose
  formalism},'' \href{http://dx.doi.org/10.1007/JHEP02(2024)171}{{\em JHEP}
  {\bfseries 02} (2024) 171}, \href{http://arxiv.org/abs/2312.17160}{{\ttfamily
  arXiv:2312.17160 [hep-th]}}.

\bibitem{Liu:2024byr}
Y.-R. Liu, J.-R. Zhang, and Y.-L. Zhang, ``{Slowly rotating charges from Weyl
  double copy for Kerr black hole with Chern\textendash{}Simons correction},''
  \href{http://dx.doi.org/10.1088/1572-9494/ad4a37}{{\em Commun. Theor. Phys.}
  {\bfseries 76} no.~8, (2024) 085405},
  \href{http://arxiv.org/abs/2404.07888}{{\ttfamily arXiv:2404.07888 [gr-qc]}}.

\bibitem{Chawla:2024mse}
S.~Chawla, K.~Fransen, and C.~Keeler, ``{The Penrose limit of the Weyl double
  copy},'' \href{http://arxiv.org/abs/2406.14601}{{\ttfamily arXiv:2406.14601
  [hep-th]}}.

\bibitem{Armstrong-Williams:2024bog}
K.~Armstrong-Williams, N.~Moynihan, and C.~D. White, ``{Deriving Weyl double
  copies with sources},'' \href{http://arxiv.org/abs/2407.18107}{{\ttfamily
  arXiv:2407.18107 [hep-th]}}.

\bibitem{Monteiro:2018xev}
R.~Monteiro, I.~Nicholson, and D.~O'Connell, ``{Spinor-helicity and the
  algebraic classification of higher-dimensional spacetimes},''
  \href{http://dx.doi.org/10.1088/1361-6382/ab03df}{{\em Class. Quant. Grav.}
  {\bfseries 36} (2019) 065006},
  \href{http://arxiv.org/abs/1809.03906}{{\ttfamily arXiv:1809.03906 [gr-qc]}}.

\bibitem{Chawla:2022ogv}
S.~Chawla and C.~Keeler, ``{Aligned fields double copy to Kerr-NUT-(A)dS},''
  \href{http://dx.doi.org/10.1007/JHEP04(2023)005}{{\em JHEP} {\bfseries 04}
  (2023) 005}, \href{http://arxiv.org/abs/2209.09275}{{\ttfamily
  arXiv:2209.09275 [hep-th]}}.

\bibitem{Didenko:2011ir}
V.~E. Didenko, ``{Coordinate independent approach to 5d black holes},''
  \href{http://dx.doi.org/10.1088/0264-9381/29/2/025009}{{\em Class. Quant.
  Grav.} {\bfseries 29} (2012) 025009},
  \href{http://arxiv.org/abs/1108.4321}{{\ttfamily arXiv:1108.4321 [hep-th]}}.

\bibitem{CarrilloGonzalez:2022mxx}
M.~Carrillo~Gonz\'alez, A.~Momeni, and J.~Rumbutis, ``{Cotton double copy for
  gravitational waves},''
  \href{http://dx.doi.org/10.1103/PhysRevD.106.025006}{{\em Phys. Rev. D}
  {\bfseries 106} no.~2, (2022) 025006},
  \href{http://arxiv.org/abs/2202.10476}{{\ttfamily arXiv:2202.10476
  [hep-th]}}.

\bibitem{CarrilloGonzalez:2022ggn}
M.~Carrillo~Gonz\'alez, W.~T. Emond, N.~Moynihan, J.~Rumbutis, and C.~D. White,
  ``{Mini-twistors and the Cotton double copy},''
  \href{http://dx.doi.org/10.1007/JHEP03(2023)177}{{\em JHEP} {\bfseries 03}
  (2023) 177}, \href{http://arxiv.org/abs/2212.04783}{{\ttfamily
  arXiv:2212.04783 [hep-th]}}.

\bibitem{Ortaggio:2023rzp}
M.~Ortaggio and A.~Srinivasan, ``{Charging Kerr-Schild spacetimes in higher
  dimensions},'' \href{http://dx.doi.org/10.1103/PhysRevD.110.044035}{{\em
  Phys. Rev. D} {\bfseries 110} no.~4, (2024) 044035},
  \href{http://arxiv.org/abs/2309.02900}{{\ttfamily arXiv:2309.02900 [gr-qc]}}.

\bibitem{Ortaggio:2023cdz}
M.~Ortaggio, V.~Pravda, and A.~Pravdova, ``{Kerr-Schild double copy for Kundt
  spacetimes of any dimension},''
  \href{http://dx.doi.org/10.1007/JHEP02(2024)069}{{\em JHEP} {\bfseries 02}
  (2024) 069}, \href{http://arxiv.org/abs/2312.00706}{{\ttfamily
  arXiv:2312.00706 [gr-qc]}}.

\bibitem{Coley:2004jv}
A.~Coley, R.~Milson, V.~Pravda, and A.~Pravdova, ``{Classification of the Weyl
  tensor in higher dimensions},''
  \href{http://dx.doi.org/10.1088/0264-9381/21/7/L01}{{\em Class. Quant. Grav.}
  {\bfseries 21} (2004) L35--L42},
  \href{http://arxiv.org/abs/gr-qc/0401008}{{\ttfamily arXiv:gr-qc/0401008}}.

\bibitem{Milson:2004jx}
R.~Milson, A.~Coley, V.~Pravda, and A.~Pravdova, ``{Alignment and algebraically
  special tensors in Lorentzian geometry},''
  \href{http://dx.doi.org/10.1142/S0219887805000491}{{\em Int. J. Geom. Meth.
  Mod. Phys.} {\bfseries 2} (2005) 41--61},
  \href{http://arxiv.org/abs/gr-qc/0401010}{{\ttfamily arXiv:gr-qc/0401010}}.

\bibitem{Pravda:2004ka}
V.~Pravda, A.~Pravdova, A.~Coley, and R.~Milson, ``{Bianchi identities in
  higher dimensions},''
  \href{http://dx.doi.org/10.1088/0264-9381/24/6/C01}{{\em Class. Quant. Grav.}
  {\bfseries 21} (2004) 2873--2898},
  \href{http://arxiv.org/abs/gr-qc/0401013}{{\ttfamily arXiv:gr-qc/0401013}}.
  [Erratum: Class.Quant.Grav. 24, 1691 (2007)].

\bibitem{Ortaggio:2012jd}
M.~Ortaggio, V.~Pravda, and A.~Pravdova, ``{Algebraic classification of higher
  dimensional spacetimes based on null alignment},''
  \href{http://dx.doi.org/10.1088/0264-9381/30/1/013001}{{\em Class. Quant.
  Grav.} {\bfseries 30} (2013) 013001},
  \href{http://arxiv.org/abs/1211.7289}{{\ttfamily arXiv:1211.7289 [gr-qc]}}.

\bibitem{Newman:1961qr}
E.~Newman and R.~Penrose, ``{An Approach to gravitational radiation by a method
  of spin coefficients},''
\href{http://dx.doi.org/10.1063/1.1724257}{{\em J. Math. Phys.} {\bfseries 3}
  (1962) 566--578}.
%%CITATION = JMAPA,3,566;%%.

\bibitem{Durkee:2009nm}
M.~Durkee and H.~S. Reall, ``{A Higher-dimensional generalization of the
  geodesic part of the Goldberg-Sachs theorem},''
  \href{http://dx.doi.org/10.1088/0264-9381/26/24/245005}{{\em Class. Quant.
  Grav.} {\bfseries 26} (2009) 245005},
  \href{http://arxiv.org/abs/0908.2771}{{\ttfamily arXiv:0908.2771 [gr-qc]}}.

\bibitem{Ortaggio:2012hc}
M.~Ortaggio, V.~Pravda, A.~Pravdova, and H.~S. Reall, ``{On a five-dimensional
  version of the Goldberg-Sachs theorem},''
  \href{http://dx.doi.org/10.1088/0264-9381/29/20/205002}{{\em Class. Quant.
  Grav.} {\bfseries 29} (2012) 205002},
  \href{http://arxiv.org/abs/1205.1119}{{\ttfamily arXiv:1205.1119 [gr-qc]}}.

\bibitem{Coley:2004hu}
A.~Coley, R.~Milson, V.~Pravda, and A.~Pravdova, ``{Vanishing scalar invariant
  spacetimes in higher dimensions},''
  \href{http://dx.doi.org/10.1088/0264-9381/21/23/014}{{\em Class. Quant.
  Grav.} {\bfseries 21} (2004) 5519--5542},
  \href{http://arxiv.org/abs/gr-qc/0410070}{{\ttfamily arXiv:gr-qc/0410070}}.

\bibitem{Ortaggio:2007eg}
M.~Ortaggio, V.~Pravda, and A.~Pravdova, ``{Ricci identities in higher
  dimensions},'' \href{http://dx.doi.org/10.1088/0264-9381/24/6/018}{{\em
  Class. Quant. Grav.} {\bfseries 24} (2007) 1657--1664},
  \href{http://arxiv.org/abs/gr-qc/0701150}{{\ttfamily arXiv:gr-qc/0701150}}.

\bibitem{Coley:2002ku}
A.~Coley, R.~Milson, N.~Pelavas, V.~Pravda, A.~Pravdova, and R.~Zalaletdinov,
  ``{Generalized pp wave space-times in higher dimensions},''
  \href{http://dx.doi.org/10.1103/PhysRevD.67.104020}{{\em Phys. Rev. D}
  {\bfseries 67} (2003) 104020},
  \href{http://arxiv.org/abs/gr-qc/0212063}{{\ttfamily arXiv:gr-qc/0212063}}.

\bibitem{Coley:2006fr}
A.~Coley, A.~Fuster, S.~Hervik, and N.~Pelavas, ``{Higher dimensional VSI
  spacetimes},'' \href{http://dx.doi.org/10.1088/0264-9381/23/24/014}{{\em
  Class. Quant. Grav.} {\bfseries 23} (2006) 7431--7444},
  \href{http://arxiv.org/abs/gr-qc/0611019}{{\ttfamily arXiv:gr-qc/0611019}}.

\bibitem{Zhao:2024wtn}
W.~Zhao, P.-J. Mao, and J.-B. Wu, ``{Weyl double copy in type D spacetime in
  four and five dimensions},''
  \href{http://dx.doi.org/10.1103/PhysRevD.111.066005}{{\em Phys. Rev. D}
  {\bfseries 111} no.~6, (2025) 066005},
  \href{http://arxiv.org/abs/2411.04774}{{\ttfamily arXiv:2411.04774
  [hep-th]}}.

\bibitem{Gibbons:2004uw}
G.~W. Gibbons, H.~Lu, D.~N. Page, and C.~N. Pope, ``{The General Kerr-de Sitter
  metrics in all dimensions},''
  \href{http://dx.doi.org/10.1016/j.geomphys.2004.05.001}{{\em J. Geom. Phys.}
  {\bfseries 53} (2005) 49--73},
  \href{http://arxiv.org/abs/hep-th/0404008}{{\ttfamily arXiv:hep-th/0404008}}.

\bibitem{Gibbons:2004js}
G.~W. Gibbons, H.~Lu, D.~N. Page, and C.~N. Pope, ``{Rotating black holes in
  higher dimensions with a cosmological constant},''
  \href{http://dx.doi.org/10.1103/PhysRevLett.93.171102}{{\em Phys. Rev. Lett.}
  {\bfseries 93} (2004) 171102},
  \href{http://arxiv.org/abs/hep-th/0409155}{{\ttfamily arXiv:hep-th/0409155}}.

\end{thebibliography}

\end{document}